# Retrieval of very large numbers of items in the Web of Science: an exercise to develop accurate search strategies.

Ricardo Arencibia-Jorge,[a,c] Loet Leydesdorff,[b] Zaida Chinchilla-Rodríguez,[c] Ronald Rousseau,[d,e] and Soren W. Paris[f]

[a] *National Scientific Research Center, Avenue 25 and 158 street, AP 6414 Havana, Cuba*
[b] *Amsterdam School of Communication Research, University of Amsterdam, Kloveniersburgwal 48, 1012 CX Amsterdam, The Netherlands*
[c] *CSIC, Institute of Public Goods and Policies, SCImago Research Group, Albazans 26-28, Madrid, Spain*
[d] *K.U.Leuven, Department of Mathematics, Celestijnenlaan 200B, 3001 Leuven (Heverlee), Belgium*
[e] *K.U.Leuven, Steunpunt O&O Indicatoresn, Dekenstraat 2, B-3000, Belgium*
[f] *Thomson Reuters, 3501 Market Street, Philadelphia, PA 19104, U.S.A.*

**Abstract**

The current communication presents a simple exercise with the aim of solving a singular problem: the retrieval of extremely large amounts of items in the Web of Science interface. As it is known, Web of Science interface allows a user to obtain at most 100,000 items from a single query. But what about queries that achieve a result of more than 100,000 items? The exercise developed one possible way to achieve this objective. The case study is the retrieval of the entire scientific production from the United States in a specific year. Different sections of items were retrieved using the field *Source* of the database. Then, a simple Boolean statement was created with the aim of eliminating overlapping and to improve the accuracy of the search strategy. The importance of team work in the development of advanced search strategies was noted.

**Keywords:** Information retrieval, search strategies, databases, Web of Science.

*Address for correspondence:*
Ricardo Arencibia-Jorge
ricardo.arencibia@cnic.edu.cu
National Scientific Research Center
Avenue 25 and 158 street, AP 6414 Havana, Cuba.

# 1. Introduction

Sometimes information professionals must face singular problems related to the use of information technologies and the management of digital environments. Specially, the continuous changes and improvements experimented by the online interfaces of well-known information resources, which also require a constant learning activity in their daily work (Martínez, 2008). There is always a moment in which a little an easy activity depends on the practical knowledge of the specialists, their skills to share experiences with other professionals, and their abilities to work in group with a clear objective: to find the best possible solution.

This exercise came up during a work session of the SCImago research team. Specialists from the Spanish group were doing a scientometric study of the world scientific production in 2007 using Scopus and Web of Science (WoS) interfaces, when they noticed an inconvenience which at first sight was apparently easy to solve: the necessity to create search strategies to retrieve the scientific production of the USA and the United Kingdom in the WoS interface.

Recently, one of the multiple papers of Péter Jacsó on search strategies and techniques in the most widely used citation-enhanced databases, called attention to this topic (Jacsó, 2009). WoS limits the search sets to 100,000 records. So, big geopolitical domains with large amounts of items must be searched using a combination of search statements. The clearest examples were countries such as the USA or the United Kingdom, or blocks of countries such as the European Union, with a scientific production in mainstream journals of over 100,000 articles during a year.

The identification of items from the United Kingdom doesn't represent major difficulties. The construction of two statements including and excluding the word "London" in the field *Author Address* can easily solve the problem for the moment. For example, using all databases comprised by WoS (SCI-EXPANDED, SSCI, A&HCI, CPCI-S, CPCI-SSH, IC, CCR-EXPANDED), and selecting all years in Timespan (this is important, because items from 2007 processed the year after or before by WoS are omitted if 2007 is used in the *Limits* featured in the WoS interface), a user can obtain the total output of this nation through the sum of the items retrieved by the following search statements:

1. PY=2007 AND CU=(England OR Scotland OR Wales OR North Ireland) AND AD=LONDON
2. PY=2007 AND CU=(England OR Scotland OR Wales OR North Ireland) NOT AD=LONDON

Therefore, as of June 18 of 2009 (the date of this query), there were 33,043 articles in 2007 signed by authors from at least one London scientific or scholarly institution, and there were 98,802 in which there was no author from this English city. A total of 131,845 articles compose the sum total output of the United Kingdom in the WoS that year.

But, what about the USA? The scientific production from this country in a year far exceeds 100,000 articles. How to find the way to obtain the total output of the USA using the WoS interface? That question gave rise to a practical and interesting exercise, which required the union of various specialists from different research groups.

## 2. In search of a solution

At first, a series of search strategies developed by the SCImago group were oriented towards the identification of the states of the Union in the *Author Address* (AD) field, with the aim of obtaining different sections of less than 100,000 items. But the design of this kind of advanced search strategies, based on the AD field, were very complex in this case. The wide collaboration between institutions from different states made it very difficult to elaborate a logical operation in the search strategy so as to eliminate duplicates. Ronald Rousseau devised the most complete strategy, but the results required a very complex process of validation. All the strategies and results were sent to Eugene Garfield and his assistant Soren W. Paris from Thomson Reuters, which validated the results with their own results obtained from their direct searches in ISI databases. In this case, there were still significant differences between the AD-based search strategy and the statistics compiled by Garfield and Paris.

Loet Leydesdorff, based on previous personal experiences, proposed the use of a less problematical field to elaborate the search strategy: the *Source* (SO) field (Zhou & Leydesdorff, 2006). Thus, using the initial of the journal/proceeding title plus an asterisk (a truncation oriented to retrieve all titles with the initial selected), the process of division in sections of less than 100,000 items was effortless. The only problem was the existence of journals belonging to series, which were retrieved not only by the journal title, but also by the series title. In any case, there were only two possibilities to obtain duplicated data; that is, a journal could be covered by no more than two sections of less than 100,000 items. Therefore, the possibility to create a Boolean statement in the search strategy with the aim to eliminate duplicates was quite easy.

Leydesdorff´s proposal, improved on by the participants in the exercise, was developed by the SCImago research group, which finally elaborated a more accurate search strategy and developed the validation procedure.

**3. Proposed search strategy**

Table 1 shows the complete procedure elaborated to obtain the total number of articles produced by institutions from the United States.

**Table 1.** Search strategy to obtain the total number of articles from the United States of America in 2007 through the WoS interface (Databases=SCI-EXPANDED, SSCI, A&HCI, CPCI-S, CPCI-SSH; Timespan=All Years; All kind of documents)

| Search Strategy | Items | Sum |
|---|---|---|
| 1. PY=2007 AND CU=USA AND (SO=A* OR SO=B*) | 91 122 | *91 122* |
| 2. PY=2007 AND CU=USA AND (SO=C* OR SO=D* OR SO=E* OR SO=F* OR SO=G*) | 91 920 | *183 042* |
| 3. PY=2007 AND CU=USA AND (SO=H* OR SO=I* OR SO=K* OR SO=L* OR SO=M*) | 82 897 | *265 939* |
| 4. PY=2007 AND CU=USA AND (SO=N* OR SO=O* OR SO=P* OR SO=Q* OR SO=R*) | 84 783 | *350 722* |
| 5. PY=2007 AND CU=USA AND (SO=S* OR SO=T* OR SO=U* OR SO=V* OR SO=W* OR SO=X* OR SO=Y* OR SO=Z* OR SO=1* OR SO=2* OR SO=3* OR SO=4* OR SO=5* OR SO=6* OR SO=7* OR SO=8* OR SO=9*) | 58 751 | *409 473* |
| 6. PY=2007 AND CU=USA AND SO=J* AND AD=CA | 17 064 | *426 537* |
| 7. PY=2007 AND CU=USA AND SO=J* NOT AD=CA | 92 976 | ***519 513*** |
| **Statement to find overlapping** | **Items** | **Sum - Overlapping** |
| 8. (#1 AND #2) OR (#1 AND #3) OR (#1 AND #4) OR (#1 AND #5) OR (#1 AND #6) OR (#1 AND #7) OR (#2 AND #3) OR (#2 AND #4) OR (#2 AND #5) OR (#2 AND #6) OR (#2 AND #7) OR (#3 AND #4) OR (#3 AND #5) OR (#3 AND #6) OR (#3 AND #7) OR (#4 AND #5) OR (#4 AND #6) OR (#4 AND #7) OR (#5 AND #6) OR (#5 AND #7) OR (#6 AND #7) | **23 026** (Overlapping) | **496 487** (Σ 1-7) - 8 |
| **New Search Strategy (Excluding overlapping)** | **Items** | **Sum** |
| 9. #1 NOT #8 | 85 586 | *85 586* |
| 10. #2 NOT #8 | 87 535 | *173121* |
| 11. #3 NOT #8 | 69 457 | *242578* |
| 12. #4 NOT #8 | 75 516 | *318094* |
| 13. #5 NOT #8 | 45 551 | *363645* |
| 14. #6 NOT #8 | 17 008 | *380653* |
| 15. #7 NOT #8 | 92 808 | ***473 461*** |
| **Sum 9-15 plus articles excluded by overlapping** | | **496 487** (Σ 9-15) + 8 |

\* Search developed in May 18, 2009.

The first 7 statements were created with the aim of dividing the results in sections of less than 100,000 items. In each statement, following alphabetical order, the necessary journal initials to obtain an upper limit of less than 100,000 results were used. Note that statements #6 and #7 were shaped with the same philosophy as the United Kingdom output retrieval procedure. There were more than 100,000 USA articles published in journals whose titles beging with "*J* ". Therefore, the AD field was used to divide this specific section in two: articles published in these journals including authors belonging to institutions from California (CA), and excluding them. In the end, a total number of 519,513 articles was obtained.

Then, a simple Boolean statement (# 8) was created in order to identify overlapping and to improve the accuracy of the search strategy. Resting these 23,026 overlapped items from the previously calculated number, a final number of 489,487 items was obtained.

With the purpose of identifying inaccuracies in the calculation process, the first 7 statements were implemented again (#9 to #15), but excluding in each case the overlapped items, which gave the result of 473,461 items. The overlapped items were added, and 489,487 items were once again obtained. So at that moment, this number comprised the hypothetical total number of articles published by institutions from the USA during the year 2007.

## 4. Validation process

The validation process was simple. First, it was taken into account the total scientific output of a less productive country than the USA or the United Kingdom was taken into account; of course: this could be any nation from the rest of the world. In this case, Cuba was used as example.

First, a direct method was used to find the Cuban scientific production in WoS during the year 2007 (Table 2).

**Table 2.** Search strategy to obtain the total number of Cuban articles in 2007 through the WoS interface: direct method (Databases=SCI-EXPANDED, SSCI, A&HCI, CPCI-S, CPCI-SSH; Timespan=All Years; All kind of documents)

| Search Strategy | Items | Sum |
|---|---|---|
| 1. PY=2007 AND CU=CUBA | 910 | **910** |

A total of 910 items were identified using the word "Cuba" in the *Affiliation Country* (CU) field. So, the second step was to use the same strategy developed to retrieve the total USA output. If the search strategy was correctly developed, the final number to obtain by either of the two indirect methods (including and excluding overlapped items) had to be 910, neither more nor less. Table 3 confirms, finally, the accuracy of data obtained through the search strategy developed during the exercise.

**Table 3.** Search strategy to obtain the total number of Cuban articles in 2007 through the WoS interface: indirect methods (Databases=SCI-EXPANDED, SSCI, A&HCI, CPCI-S, CPCI-SSH; Timespan=All Years; All kind of documents)

| Search Strategy | Items | Sum |
|---|---|---|
| 1. PY=2007 AND CU=CUBA AND (SO=A* OR SO=B*) | 140 | *140* |
| 2. PY=2007 AND CU=CUBA AND (SO=C* OR SO=D* OR SO=E* OR SO=F* OR SO=G*) | 216 | *356* |
| 3. PY=2007 AND CU=CUBA AND (SO=H* OR SO=I* OR SO=K* OR SO=L* OR SO=M*) | 161 | *517* |
| 4. PY=2007 AND CU=CUBA AND (SO=N* OR SO=O* OR SO=P* OR SO=Q* OR SO=R*) | 193 | *710* |
| 5. PY=2007 AND CU=CUBA AND (SO=S* OR SO=T* OR SO=U* OR SO=V* OR SO=W* OR SO=X* OR SO=Y* OR SO=Z* OR SO=1* OR SO=2* OR SO=3* OR SO=4* OR SO=5* OR SO=6* OR SO=7* OR SO=8* OR SO=9*) | 91 | *801* |
| 6. PY=2007 AND CU=CUBA AND SO=J* AND AD=Havana | 108 | *909* |
| 7. PY=2007 AND CU=CUBA AND SO=J* NOT AD=Havana | 35 | ***944*** |
| **Statement to find overlapping** | **Items** | **Sum - Overlapping** |
| 8. (#1 AND #2) OR (#1 AND #3) OR (#1 AND #4) OR (#1 AND #5) OR (#1 AND #6) OR (#1 AND #7) OR (#2 AND #3) OR (#2 AND #4) OR (#2 AND #5) OR (#2 AND #6) OR (#2 AND #7) OR (#3 AND #4) OR (#3 AND #5) OR (#3 AND #6) OR (#3 AND #7) OR (#4 AND #5) OR (#4 AND #6) OR (#4 AND #7) OR (#5 AND #6) OR (#5 AND #7) OR (#6 AND #7) | **34** (Overlapping) | **910** (Σ 1-7) - 8 |
| **New Search Strategy (Excluding overlapping)** | **Items** | **Sum** |
| 9. #1 NOT #8 | 127 | *127* |
| 10. #2 NOT #8 | 205 | *332* |
| 11. #3 NOT #8 | 139 | *471* |
| 12. #4 NOT #8 | 177 | *648* |
| 13. #5 NOT #8 | 86 | *734* |
| 14. #6 NOT #8 | 108 | *842* |
| 15. #7 NOT #8 | 34 | ***876*** |
| **Sum 9-15 plus articles excluded by overlapping** | | **910** (Σ 9-15) + 8 |

\* Search developed in May 19, 2009.

In the end, data obtained in WoS following this search strategy had a total coincidence with data reported by the Thomson Reuters team.

## 5. Final considerations

The exercise allows us to obtain the same in two different ways: a) searching with overlapping, and subtracting items overlapped at the end; and b) searching without overlapping and adding the items overlapped at the end. The validation procedure in a small country also allows us to obtain the same total number not only through the strategies proposed, but also using a direct method, which confirms the accuracy of the results and the efficacy of the statements designed.

Probably this kind of SO-based search strategy is not the unique alternative to retrieve the USA scientific production in WoS, and it may be that its implementation does not solve other problems related to large amounts of items to be retrieved through the WoS interface. It may even be the case that somebody has developed a similar idea before, as a result of a logical reflection in their daily work.

In any case, for scientometric purposes, a fast and well described method to obtain reliable data is always welcome. In this sense, the current proposal is an accurate and validated search strategy to be used by any specialist around the world, and the procedure presented shows the importance of team work in the development of advanced search strategies for information retrieval.

**Acknowledgments**

To Félix de Moya Anegón, Carmen López-Illescas, Elena Corera Álvarez, and María Benavent Pérez (SCImago Research Group, Institute of Public Goods and Policies CSIC), for their support and advices. Thomson Reuters databases were available in Spain thanks to the Spanish Foundation for Science and Technology and the Ministry of Science and Innovation of the Spanish government.

Martínez, L.J. (2008). La nueva versión de ISI Web of Knowledge: calidad y complejidad. *El Profesional de la Información*, 17, 331-339.

Zhou, P., & Leydesdorff, L. (2006). The emergence of China as a leading nation in science. *Research Policy,* 35, 83-104.
8